\documentclass[
    ,final            
  ]
  {aipproc}

\layoutstyle{6x9}

\begin{document}

\title{A New Godunov Scheme for MHD, with Application to the MRI in disks}

\classification{95.30.Qd; 97.10.Gz}
\keywords      {MHD; accretion disks}

\author{James M. Stone}{
  address={Department of Astrophysical Sciences, Princeton University, Princeton, NJ 08544}
}

\author{Thomas A. Gardiner}{
  address={Department of Astrophysical Sciences, Princeton University, Princeton, NJ 08544}
}


\begin{abstract}
We describe a new numerical scheme for MHD which combines a higher
order Godunov method (PPM) with Constrained Transport.  The results from a
selection of multidimensional test problems are presented.  The complete
test suite used to validate the method, as well as implementations of
the algorithm in both F90 and C, are available from the web.  A fully
three-dimensional version of the algorithm has been developed, and is
being applied to a variety of astrophysical problems including the
decay of supersonic MHD turbulence, the nonlinear evolution of the MHD
Rayleigh-Taylor instability, and the saturation of the magnetorotational
instability in the shearing box.  Our new simulations of the MRI represent
the first time that a higher-order Godunov scheme has been applied to this
problem, providing a quantitative check on the accuracy of previous
results computed with ZEUS; the latter are found to be reliable. 

\end{abstract}

\maketitle


\section{Introduction}

Numerical simulations have emerged as a powerful tool for the study of
nonlinear and time-dependent magnetohydrodynamic (MHD) flows.
One particularly fruitful application has been to the
study of the saturation of the magnetorotational instability (MRI)
in accretion disks.  Numerical simulations which use a local, linear
expansion of the equations of motion, termed the {\em shearing box}
\cite{HGB}, have been used to follow the growth and saturation of
the MRI over
hundreds of orbits, starting from a variety of initial field configurations and
strengths.  This has allowed careful study of the saturation mechanism
in ideal MHD (Sano, these proceedings), as
well as the effect of a variety of additional physics on
the nonlinear regime: for example, non-ideal MHD
effects in protoplanetary disks \cite{SanoStone},
saturation in radiation dominated disks \cite{Turner},
and kinetic effects in nearly collisionless plasmas \cite{Sharma}.

In addition to local studies using the shearing box, a variety of authors
have begun to study the {\em global} dynamics of accretion flows driven by the
MRI.  For example, fully three-dimensional general relativistic MHD simulations
of the evolution of a weakly magnetized torus in the Kerr metric
have been presented by \cite{HawleyDeVilliers}, among others.  
Other global studies have been used
to investigate the interaction of a magnetized disk with a central star
\cite{Lovelace}, and
the formation of time-dependent jets and outflows (Shibata, these
proceedings), to name just a few examples.

Of course, all such studies are predicated on the availability
of accurate and robust numerical algorithms for MHD.  Much of the
work on the MRI has been based on the algorithms in the ZEUS code
(\cite{SNa}, \cite{SNb}), the simplicity of which 
has allowed the diverse physics discussed above (non ideal MHD, GRMHD,
radiation MHD)
to be added more easily.  However, there are some accretion flow problems
for which the algorithms in ZEUS may not be optimal.  For example,
global studies of geometrically thin disks (where $H/R \ll 1$,
$H$ is the vertical scale height in the disk) which span many tens of $H$ in
radius with sufficient resolution per scale height to resolve the inertial
range of turbulence driven by the MRI (say 128 grid points per $H$) would
require intractably large grids if uniform zoning is used.  Instead, such
calculations would be more feasible if nested grids could be used, so
that the finest resolution grid is confined to a few $H$ near the midplane,
and progressively coarser grids are used to span the less dense corona
above the disk (where the fields become strong, the MRI is suppressed, and
turbulence is reduced \cite{MillerStone}).
However, nested and adaptive grids work
best with single-step Eulerian methods based on the conservative form.
Thus, to enable new nested grid simulations of
accretion flows which span a large range in spatial and temporal scales,
we are motivated to implement a
new MHD algorithm based on the conservative form.

\section{The Basic Algorithm}

The equations of ideal MHD can be written in
conservative form as
\begin{eqnarray}
\frac{\partial \rho}{\partial t} +
{\bf\nabla\cdot} \left(\rho{\bf v}\right) & = & 0
\label{eq:cons_mass} \\
\frac{\partial \rho {\bf v}}{\partial t} +
{\bf\nabla\cdot} \left(\rho{\bf vv} - {\bf BB}\right) +
{\bf \nabla} P^* & = & 0 \\
\frac{\partial {\bf B}}{\partial t} +
{\bf\nabla\cdot} \left({\bf v B - B v}\right) & = & 0 \\
\frac{\partial E}{\partial t} +
\nabla\cdot((E + P^*) {\bf v} - {\bf B} ({\bf B \cdot v})) & = & 0
\label{eq:cons_energy}
\end{eqnarray}
where $\rho$ is the mass density, $\rho{\bf v}$ is the momentum
density, ${\bf B}$ is the magnetic field, and $E$ is the total energy
density. The total pressure $P^* \equiv P + ({\bf B \cdot B})/2$ where $P$
is the gas pressure, and the total energy density $E$ is related to the
internal energy density $\epsilon$ via
\begin{equation}
E \equiv \epsilon + \rho({\bf v \cdot v})/2 + ({\bf B \cdot B})/2 ~.
\end{equation}
Throughout we assume an ideal gas equation of state
for which $P = (\gamma - 1) \epsilon$, where $\gamma$ is the ratio of
specific heats.  Unless otherwise stated, we take $\gamma=5/3$.  None
of the main results described here depend directly upon the
equation of state.  Note also that we have chosen a system of units in
which the magnetic permeability $\mu=1$.

A variety of numerical methods can be used to solve the equations
of MHD in conservative form.  We have chosen to adopt an algorithm
based on Godunov's method (for shock capturing),
and the constrained transport
(CT) \cite{Evans-Hawley} algorithm for integration of the magnetic fields
(so that the divergence free constraint is enforced).
There are many Godunov methods we could potentially adopt;
we have chosen to use
the piecewise parabolic method (PPM, \cite{ColellaWoodward}).
A disadvantage of PPM is that it requires
the use of a MHD Riemann solver to compute the fluxes of the
volume-averaged variables (we use Roe's linearization to construct
our solver, see \cite{Cargo-Gallice}).  However, the benefit is
a very accurate method for problems involving shocks and discontinuities,
which are very prevalent in astrophysical flows.

\begin{figure}
  \includegraphics[height=.3\textheight]{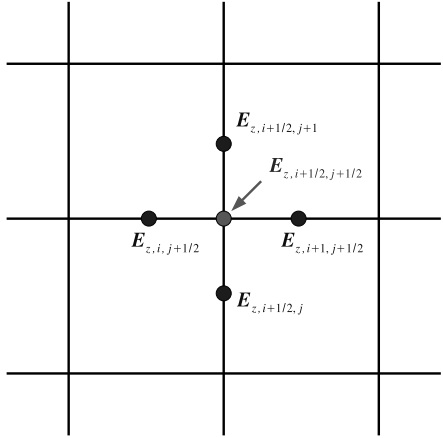}
  \caption{Schematic diagram showing location of fluxes of
volume-averaged magnetic field (cell faces) and area-averaged magnetic
field (cell corner).  The central issue in combining CT with a Godunov
scheme is developing the appropriate mapping between these fluxes.}
\end{figure}

A variety of other authors
\cite{Balsara-Spicer,Crockett,Dai-Woodward-97,Falle98,Londrillo-DelZanna03,Pen-MHD-03,RMJF98},
have described the combination CT with Godunov schemes (including PPM),
however the scheme used here differs in several important respects:
(1) in the reconstruction of cell-centered quantities to cell edges to
serve as the initial states for a Riemann solver, (2) in the way the
electromotive forces (EMFs) at cell corners needed by the CT algorithm
are computed from face-centered fluxes, and (3) in the extension
of an unsplit integration algorithm due to \cite{Colella-CTU} (the
corner-transport-upwind, or CTU, scheme) to MHD.

It is impossible to
describe all of the details of the algorithm we have developed in this
proceeding; a comprehensive discussion of the two-dimensional
algorithm is given in \cite{GS05}. However, the basic design
principle of the method is worth
reiterating here:  the relationship
between the cell-centered, volume-averaged variables used in PPM, and
the face-centered, area-averaged field components used by CT must be used to
define how the fluxes of the volume-averaged variables
(returned by the Riemann solver) are transformed into fluxes of the
area-averaged variables (needed by CT).
The issue is illustrated in Figure 1, which shows the relative
locations of the fluxes of the volume-averaged magnetic field
returned by the Riemann solver at the cell faces, and the
flux of the area-averaged magnetic field at the cell corner
($E_{z,i+1/2,j+1/2}$) used by CT.  It is natural to try using an
arithmetic average to compute the corner-centered EMF, however this
destroys the proper upwinding of the fluxes, and leads to oscillatory
solutions in many cases.  This is easy
to demonstrate for a plane parallel grid-aligned flow.  Imagine a plane
wave propagating along the $x-$direction in Figure 1.  In this case, the
corner-centered EMF should be identical to the
face-centered fluxes along the $y-$axis, that is $E_{z,i+1/2,j+1}$ and
$E_{z,i+1/2,j}$.  However, arithmetic averaging will include contributions
from the other faces, namely $E_{z,i,j+1/2}$ and $E_{z,i+1,j+1/2}$.  This
destroys the upwinding of the fluxes, and reduces the stability of
the algorithm.  Rather
than an averaging, what is needed is a reconstruction of the EMFs to the
corner; one method to achieve this is detailed in \cite{GS05}.

\section{Some Tests and First Applications}

A variety of papers discuss collections of test problems for MHD codes,
for example \cite{Toth}.  Similarly, we have put together a collection
of tests that have been useful in developing our methods.
A complete description of our test suite is given at
\texttt{http://www.astro.princeton.edu/$\sim$jstone/tests} (at the time
this article was written).
Below, we describe the results from a selection of problems.

\subsection{The Field Loop Test}

One of the simplest tests, yet challenging for a Godunov scheme, is the
advection of a field loop in multidimensions.  
In our version of this test,
the computational domain extends from $-1\le x \le 1$,
and $-0.5 \le y \le 0.5$, is resolved on a $2N \times N$ grid, and has
periodic boundary conditions on both $x$- and $y$-boundaries.
The velocity components are $v_x =
\sqrt{5} \cos(\theta)$, $v_y = \sqrt{5} \sin(\theta)$, and $v_z=0$ (where
$\cos(\theta)=2/\sqrt{5}$ and $\sin(\theta)=1/\sqrt{5}$), and $\rho=1$ and the 
pressure $P=1$.
By $t=1$ the field loop will have been advected
across the grid once.  The
$z$-component of the magnetic field $B_z=0$ while the in plane
components $B_x$ and $B_y$ are initialized from the $z$-component of
the magnetic vector potential
\begin{equation}
A_z \equiv \left \{
\begin{array}{ll}
A_0 (R - r) & \textrm{for}~ r \le R \\
0           & \textrm{for}~ r  >  R
\end{array}
\right .
\end{equation}
where $A_0=10^{-3}$, $R=0.3$ and $r=\sqrt{x^2+y^2}$.  Thus for $r \le
R$, $\beta = 2 P/B^2 = 2\times 10^{6}$ and the magnetic field is
essentially a passive scalar.

Of course, the correct solution to this test is that the field loop should
remain circular, and it should diffuse in amplitude as little as possible.
Figures of the solution computed with our method are given in \cite{GS05}
and on the test web page and will not be reproduced here; they confirm
our method preserves these properties.  We find this test provides a
dramatic demonstration of the failure of arithmetic averaging for the
construction of EMFs at cell-corners from face-centered fluxes: this
method results in a strongly oscillatory solution (although
the oscillations can be masked if a very diffusive integrator is used).
Setting the out-of-plane component of the velocity $v_z$ to a uniform, non-zero
value, and testing that $B_z$ remains zero to round-off in this case, is 
another difficult test.

\subsection{Linear Wave Convergence}

A useful quantitative test is to measure the rate at which errors
associated with the propagation of linear modes from each wave family
(entropy waves, fast and slow magnetosonic waves, and Alfven waves) converge
as the numerical resolution is increased.  Exact eigenmodes
are initialized on a three-dimensional grid of size $2L \times L
\times L$ using $2N \times N \times N$ grid points, where $N=8, 16, 32,
64$ and 128.  Thus, the grid is rectangular, but each cell is square.
We orient the wavevector such that the wavefront is along the diagonal
of the computational domain, and have chosen the grid dimensions so that
the distance between wavefronts along any coordinate direction is equal
to the size of the domain in that direction.
Periodic boundary conditions are used.
Since the wave does not propagate along the diagonal of the
grid cells, we guarantee the $x-$, $y-$, and $z-$fluxes differ;
that is the problem is truly multi-dimensional.

The background medium has $\rho = 1$,  and $P = 1/\gamma$ with $\gamma
= 5/3$.  The background magnetic field parallel to the wavevector
$b_{1} = 1$, while the two components perpendicular to it are $b_2 =
\sqrt{2}$, and $b_3 = 0.5$ (where $b = B / (4\pi)^{1/2}$).  Thus, the
fast magnetosonic speed is 2.0, the Alfven speed is 1.0, and the slow
magnetosonic speed is 0.5.  The wave is added as a perturbation to these
constant values of the form $\delta {\bf U} = A {\bf R} \sin (2 \pi l)$,
where $l$ is the displacement along the wavevector.  Here $\bf{U}$ is the
vector of conserved variables, $A$ is an amplitude, and $\bf{R}$ is the
right-eigenvector corresponding to the desired wave family.  To aid others
in comparing to our results, tables of the right-eigenvectors for the
specific initial conditions described above can be found in \cite{GS05}.
For all the tests shown here, $A = 10^{-6}$.  The face-centered components
of the magnetic field are initialized from a vector potential located
at zone edges.

After the wave has propagated one
wavelength, we measure the error in the numerical solution by computing
the norm of the vector resulting from summing the absolute value of
errors in each variable over the grid, that is we compute $\epsilon =
|| \Delta {\bf U} || = [ \sum_k (\Delta {\bf U}_k)^2 ]^{1/2}$, where
$\Delta {\bf U}_k = \sum_i | {\bf U}_{k,i}^{n} - {\bf U}_{k,i}^{0}|
/ 2N^{3}$.  Here, ${\bf U}_{k,i}^{n}$ is the numerical solution for the k-th
component of the vector of conserved quantities at grid point $i$ and
time level $n$, and ${\bf U}_{k,i}^{0}$ is the initial numerical solution
(at time zero).

\begin{figure}[t]
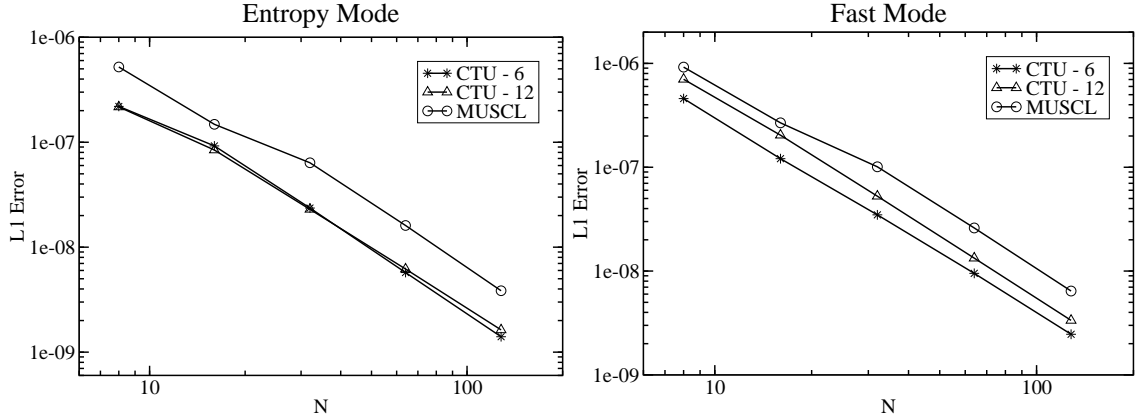

  \includegraphics[height=.25\textheight]{LinearWave-Entropy-3D.eps}
  \includegraphics[height=.25\textheight]{LinearWave-Fast-3D.eps}
  \caption{Errors in entropy and fast magnetosonic waves after one
grid crossing time computed using three different unsplit integrators.}
\end{figure}
\begin{figure}
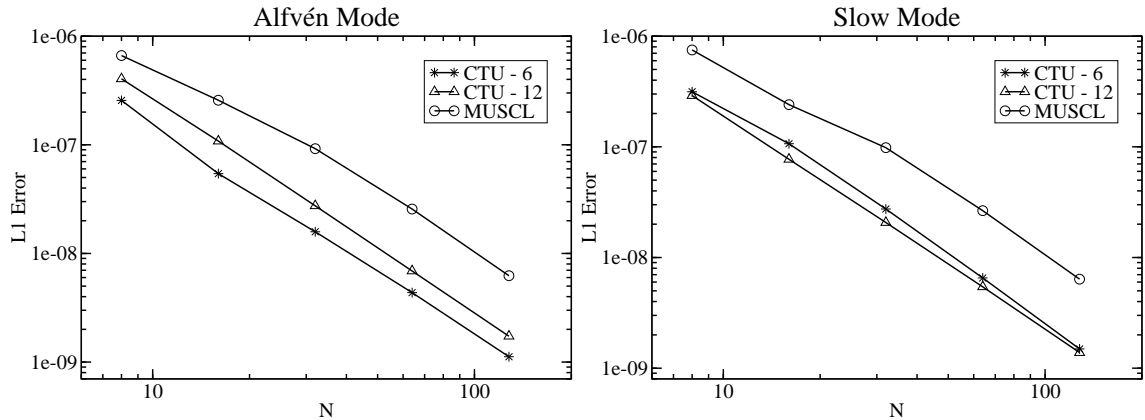

  \includegraphics[height=.25\textheight]{LinearWave-Alfven-3D.eps}
  \includegraphics[height=.25\textheight]{LinearWave-Slow-3D.eps}
  \caption{Same as Fig. 2, except for Alfven and slow magnetosonic waves.}
\end{figure}

Figures 2 and 3 plot the errors after one grid crossing time for each
wave family.  The results have been computed using three distinct
unsplit integrators: (1) the general extension of CTU to three-dimensions,
which requires 12 Riemann solves per grid cell (labeled CTU-12),
(2) a simpler extension of CTU which includes fewer transverse flux
gradients and therefore requires only 6 Riemann solves per
grid cell, but is stable up to a CFL number of only 0.5 (labeled CTU-6)
(3) a simple unsplit MUSCL scheme which is commonly
used in other codes \cite{Falle98} (labeled MUSCL).  In each case the errors converge
at second order.  However, there is a significant difference in the
amplitude of the errors at a given resolution between the integrators.
For example, the CTU-6 scheme can have an error nearly an order of magnitude
less than the MUSCL scheme for some waves and some resolutions.  This and
other tests indicate the CTU-6 scheme is the best compromise between
accuracy and simplicity; this integrator has been used for all the other
tests and application presented here.

\subsection{Nonlinear Circularly Polarized Alfven Waves}

In some sense, linear waves are too easy a test since the error
is dominated by only a few terms in the equations.
Circularly polarized Alfven waves are an
exact solution to the nonlinear equations, thus they provide
an excellent quantitative test \cite{Toth}.
For this test $\rho = 1$, $P = 0.1$,
$V_{\bot} = 0.1 \sin(2\pi x_{\|})$, 
$B_{\bot} = 0.1 \sin(2\pi x_{\|})$, and
$V_z = B_z = 0.1 \cos(2\pi x_{\|})$ with
$\gamma = 5/3$ and $x_{\|} = (x \cos\alpha + y \sin\alpha)$ where
$\alpha$ is the angle at which the wave propagates with respect to the grid.
Here $V_{\bot}$ and $B_{\bot}$ are the components of velocity
and magnetic field perpendicular to the wavevector.
Here we report on results from the two-dimensional version of this test,
using a grid of size
$2N \times N$, with the wavefront orientated along the grid diagonal
(but not the diagonal of individual zones).

Although nonlinear amplitude Alfven waves are subject to a parametric
instability which causes them to decay into magnetosonic waves (see
\cite{DelZanna}), the instability should not be present for the parameters
defined here.  Since the problem is smooth, it can be used for convergence
testing.  Running the test with smaller pressure (higher $\beta$;)
and/or larger amplitudes is a good test of how robust is the algorithm.
Using standing waves (in which the fluid is moving at the Alfven speed
in the opposite sense to the propagation of the wave) is challenging
because it requires the two terms in the EMF cancel exactly.
It is difficult to get this cancellation using schemes based on splitting

\begin{figure}
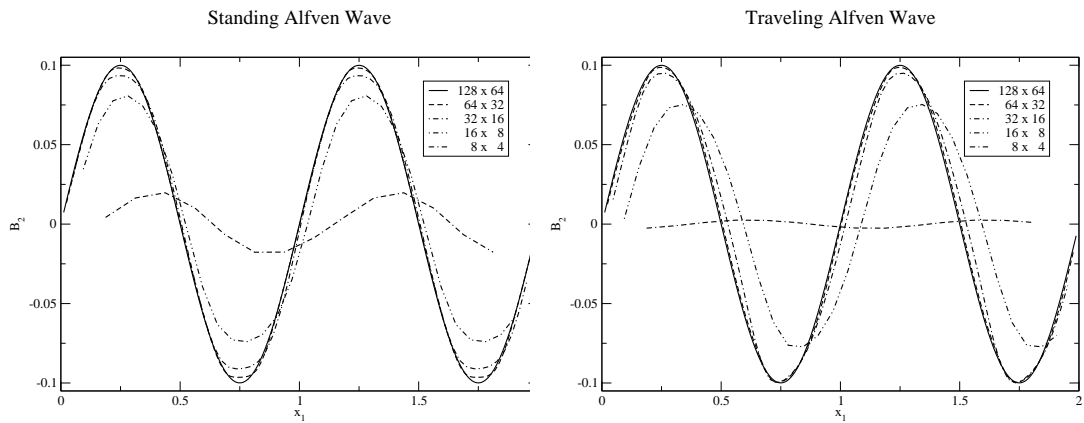

  \includegraphics[height=.25\textheight]{cp_alfven.standing.eps}
  \includegraphics[height=.25\textheight]{cp_alfven.traveling.eps}
  \caption{Profiles of standing (left) and traveling (right) circularly
polarized Alfven waves after five grid transit times at different numerical
resolutions.  All grid points in the domain are plotted; the lack
of scatter in the lines indicate there is no parametric instability}
\end{figure}

Figure 4 plots every grid point in the calculation after five grid
transit times for both standing and traveling waves.  It can be compared
directly with Figures 8 and 9 in \cite{Toth}, although note we use square grid
cells which destroys any symmetries in the fluxes (potentially
making our test more difficult).  The lack of scatter
indicates the waves have remained planar, and there is no parametric
instability.  Moreover, the profiles show very little dispersion or
diffusion error for these waves for resolutions above 16 grid points
per wavelength.  For standing waves, the wave profile is captured even
on a $8 \times 4$ grid; remarkable given that the Nyquist limit for the
minimum number of points needed to represent the wave is four.
As in the linear wave test, it is also possible to compute the convergence
rate of errors; the method here converges at second order (which is clear
from inspection of the curves in Figure 4).

\subsection{Some First Applications}

The code described above (which we call Athena),
including documentation, is freely available at
\texttt{http://www.astro.princeton.edu/$\sim$jstone}.  Currently,
one- and two-dimensional versions are available in both Fortran90 and C.
A fully three dimensional version has been developed, and is now
being used for a variety of applications, some of which we describe below.
There are significant modifications required to extend the 
method \cite{GS05} to three-dimensions, a paper is in
preparation.

One recent application of the ZEUS code was to the study of the decay
of supersonic MHD turbulence thought to dominate the internal dynamics
of cold molecular clouds.  The ZEUS simulations demonstrated that 
supersonic turbulence decays rapidly, even if it is sub-Alfvenic, because
of dissipation in slow magnetosonic shocks.  Even though the dissipation rate
in shocks could be measured in the ZEUS simulations (and was shown to
be roughly equal to the measured decay rate), it is important to test
that numerical dissipation is not strongly affecting this result.  We have
repeated the simulations in \cite{SGO} using Athena, with
the identical parameters and spectrum for the velocity driving.

The time evolution of the total energy in fluctuations (kinetic plus magnetic)
from two simulations at two different numerical resolutions, for
$\beta = 0.01$ and $\dot{E} = 1000$ (see \cite{SGO} for a detailed 
explanation of these parameters), is shown in Figure 5.  As before,
the energy shows rapid growth at early times, followed by roughly constant
evolution after 0.1 $t_s$ (where $t_s$ is the sound crossing time).
The amplitude reached by the turbulent fluctuations in the identical runs
performed with ZEUS are shown by arrows on the right axis.  In each case, the
Athena results show slightly higher amplitudes consistent with slightly
less numerical dissipation.  However, the differences are small, and the result
that the turbulence is rapidly decaying is confirmed.  We are undertaking
a survey over a much larger range in
resolutions and at different
magnetic field strengths; the results will be reported elsewhere.

\begin{figure}
  \includegraphics[height=.25\textheight]{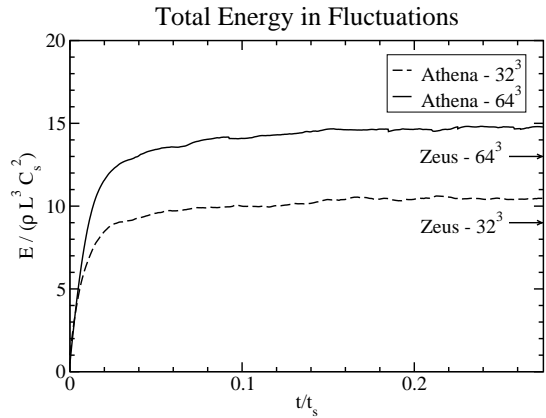}
  \caption{Time evolution of the magnetic and kinetic energy in fluctuations
for driven turbulence computed with Athena using two resolutions.  Compare
to Figure 1 of \cite{SGO}.}
\end{figure}

Another application of Athena is to nonlinear evolution of the
Rayleigh-Taylor (RT) instability in magnetized fluids.  Previous studies
using ZEUS \cite{Jun} focused on the amplification of magnetic field
in turbulence driven by the RT instability in weakly magnetized flows.
With Athena, we have computed the evolution in three-dimensions of a RT
unstable interface with strong parallel fields, in some cases including
a rotation of the field at the interface, at a grid resolution of $200
\times 200 \times 300$.  The structure of the fingers characteristic of
the RT instability at late times is quite different in the presence of
strong fields.  Since interchange instabilities at short wavelengths are
still present, whereas only long wavelength parallel modes can exist,
the fingers become ``arches" and ``flux-tubes" in the field direction.
Lack of space prevents a full discussion of the results here; a paper
is in preparation.

\section{Shearing Box Simulations of the MRI}

It is natural to compare simulations of the MRI using Athena
with the results of previous studies using the shearing box.
However, since the shearing box is a noninertial frame, source terms must be added to the equations of motion to account for the Coriolis force and tidal
gravity.  These terms,
\begin{eqnarray}
\bf{S_{\rho V_x}} & = & 3 \Omega_0^2 x \rho + 2 \Omega_0 \rho V_y \\
\bf{S_{\rho V_y}} & = & - 2 \Omega_0 \rho V_x \\
\bf{S_{E}} & = & 3 \Omega_0^2 x \rho V_x 
\end{eqnarray}
must be added to the integrator in way which preserves the second-order
convergence properties.
By redefining the total energy to include the tidal
potential, it is possible to ensure it is conserved.  Finally,
it is important that the total Coriolis force remains strictly orthogonal
to the velocity, which also requires special care.  We are preparing a
paper which describes the (non-trivial) implementation of the shearing box in Athena, as well
as presenting new results.  Here, we focus on the
comparison between ZEUS and Athena on this problem.

\begin{figure}
  \includegraphics[height=.25\textheight]{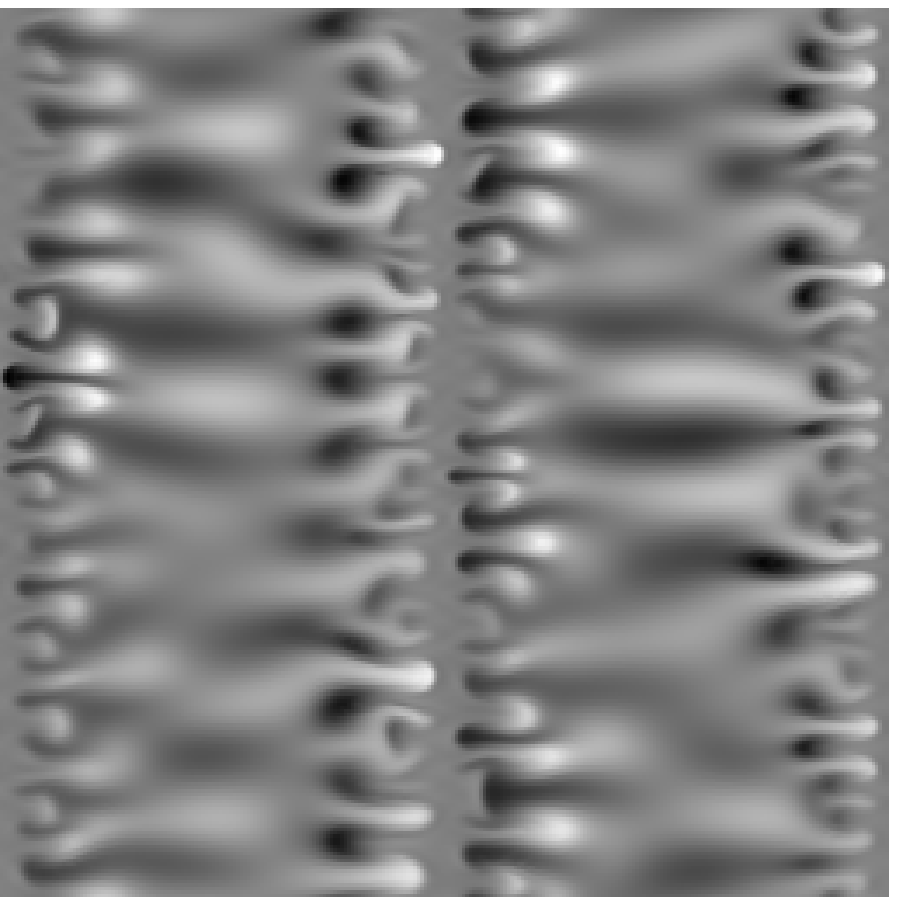}
  \includegraphics[height=.25\textheight]{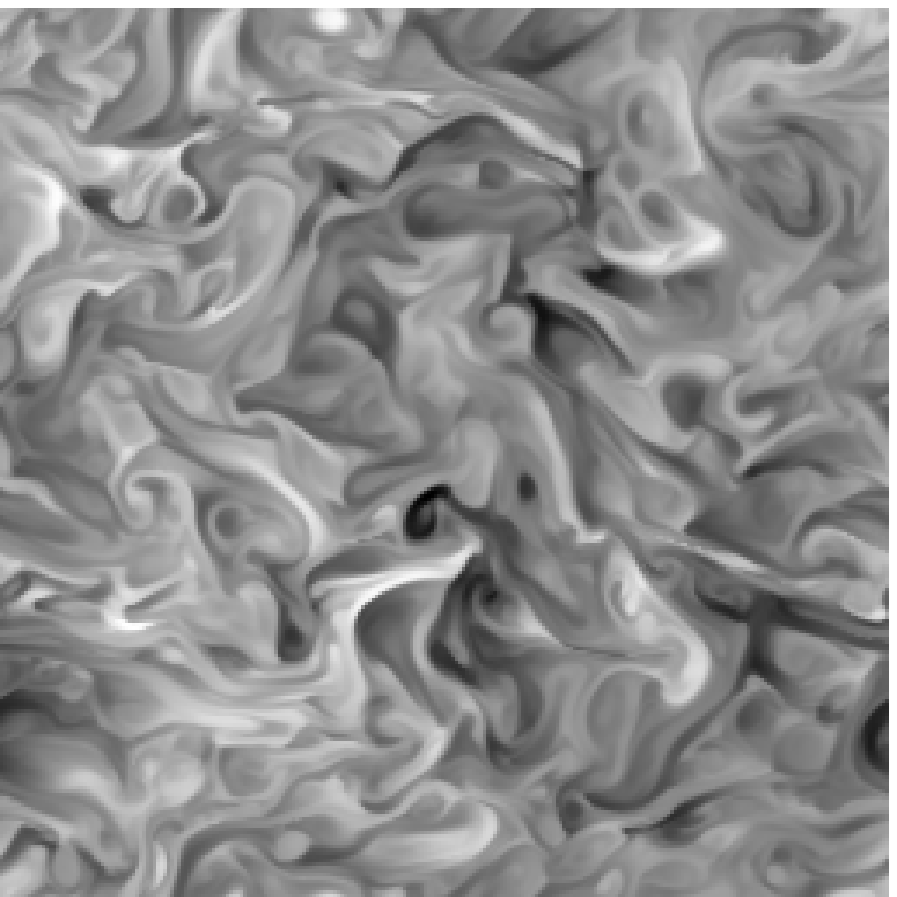}
  \caption{Velocity fluctuations at $t=3.3$ orbits (left) and $t=5.56$ orbits 
(right) for a 2D simulation of the MRI using Athena.}
\end{figure}

Figure 6 shows an image of the angular velocity fluctuations $\delta V_y
= V_y - 3 \Omega_0 x /2$ at two times in the evolution of the MRI in a
two-dimensional grid starting from a zero-net vertical field with $\beta=4000$
(identical to Run S1c in \cite{HB}, except $P_{\circ}=10^{-5}$ here) using a resolution of $256^2$.  The
panel on the left shows the growth of fingers at the characteristic wavelength
of the fastest growing mode of the MRI; since the field strength varies
with $x$, the spacing of the fingers also varies with the $x-$position.
The panel on the right shows the interaction of the fingers at late time
results in two-dimensional MHD turbulence.  Because sustained field
growth is not possible in 2D with no net flux, this turbulence inevitably dies away, with the rate of
decay determined by the numerical reconnection rate.  Thus, the time
evolution of the poloidal magnetic energy is a good measure of the numerical
dissipation rate.

\begin{figure}
  \includegraphics[height=.25\textheight]{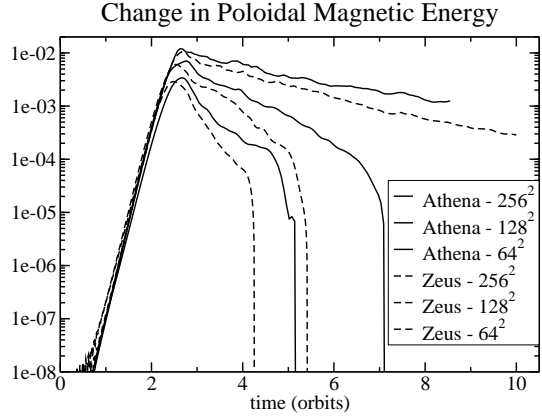}
  \caption{Evolution of the change in
poloidal magnetic energy in a 2D simulation of
the MRI starting from zero-net vertical flux.}
\end{figure}

Figure 7 shows a comparison of the evolution of the change in the volume
averaged poloidal
magnetic energy $<\delta B^{2}> = <B^{2}> - <B_{0}^{2}>$, where $<B_{0}>$
is the volume averaged magnetic energy in the initial state, at
three different resolutions computed with both ZEUS and Athena.  Note that
at each resolution, the Athena curve is about half way between the ZEUS curves
of the same and the next highest resolution.  This indicates the numerical
dissipation in Athena is similar to that in ZEUS at about 1.5 times higher
resolution per dimension.  The result that the dissipation in Athena is
slightly lower than ZEUS is consistent with the turbulence decay results
presented earlier.  Note that there are no qualitative differences between
the ZEUS and Athena results, the growth rate and saturation amplitude of
the MRI is the same in each case.  This is the first time the evolution
of the MRI has been computed with a higher order Godunov code, and is a
powerful confirmation of the previous ZEUS results.

\begin{figure}
  \includegraphics[height=.35\textheight]{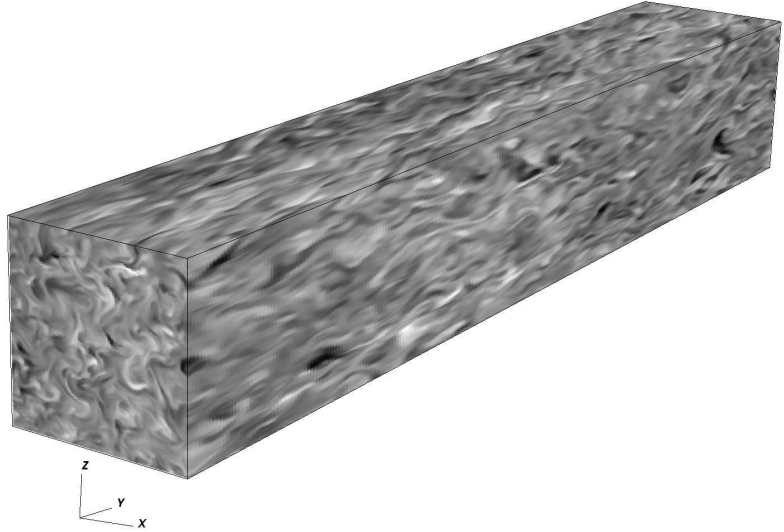}
  \caption{Angular velocity fluctuations on the surface of the
computational volume for a 3D simulation of the MRI computed with Athena.}
\end{figure}

Figure 8 shows the angular momentum fluctuations on the faces of a
three-dimensional computational volume computed using a grid of
$128 \times 256 \times 128$ and an initially zero-net vertical field of strength
given by $\beta = 4000$ (with $P_{\circ}=10^{-6}$), similar to
Run SZ1 in \cite{HGB2}.  In 3D, turbulence is sustained, and
the time-evolution of volume averaged quantities, including the Maxwell
stress, is very similar to the previous ZEUS results. 

In addition to comparing to results from ZEUS, we have been using
Athena to study the effect of optically thin cooling on the nonlinear
stage of the MRI in the shearing box.  Since Athena is conservative,
magnetic energy lost in reconnection is captured as thermal heating of the
gas.  Thus, Athena is much more suitable for studies of the energetics
of the MRI.  As discussed by Gardiner (these proceedings), we have
added optically thin cooling to Athena and are studying the statistics of
the resulting temperature fluctuations in steady-state turbulence driven by
the MRI.

Our plans for Athena include extension to curvilinear and nested
grids, and application to (1) the global MHD of geometrically thin disks
and (2) fragmentation and collapse in self-gravitating MHD turbulence.
These, and many other applications of Athena, will be reported in
future communications.


\begin{theacknowledgments}
We thank J. Hawley and P. Teuben for their significant contributions to the
work presented here.  This work was supported by NSF grant
AST-0113571.   We thank the organizers for hosting such a wonderful
meeting in such a beautiful place.
\end{theacknowledgments}

\end{document}